\documentclass{article}

\usepackage{epsfig}
\usepackage{graphicx}
\usepackage{array}
\usepackage{color}

\def\piplus        {$\pi^{+}$}
\def\pimin         {$\pi^{-}$}

\def\kplus         {K$^{+}$}

\def\kzeros        {K$^{0}_{s}$}

\def\lambdazero    {$\Lambda$}
\def\antilambda    {$\overline{\Lambda}$}
\def\ap            {$\overline{p}$}

\def\Pgp{\ifmmode{\mathrm \pi}
          \else${\mathrm \pi }$\fi}
\def\Pgpm{\ifmmode{\mathrm \pi^-}
           \else${\mathrm \pi^-}$\fi}
\def\Pgpp{\ifmmode{\mathrm \pi^+}
           \else${\mathrm \pi^+}$\fi}

\def\PK{\ifmmode{\mathrm K}
         \else${\mathrm K}$\fi}
\def\PKpm{\ifmmode{\mathrm K^{\pm}}
           \else${\mathrm K^{\pm}}$\fi}
\def\PKp{\ifmmode{\mathrm K^+}
          \else${\mathrm K^+}$\fi}
\def\PKm{\ifmmode{\mathrm K^-}
          \else${\mathrm K^-}$\fi}
\def\Pp{\ifmmode{\mathrm p}
         \else${\mathrm p}$\fi}
\def\PgL{\ifmmode{\mathrm \Lambda}
          \else${\mathrm \Lambda}$\fi}
\def\PagL{\ifmmode{\overline{\mathrm \Lambda}}
          \else${\overline{\mathrm \Lambda}}$\fi}
\def\PgOm{\ifmmode{\mathrm \Omega^-}
           \else${\mathrm \Omega^-}$\fi}

\def\munit{\ifmmode{\,\mathrm{MeV/{\mit c}^{\,2}}}
          \else{MeV/$c^{\,2}$}\fi}
\def\mup{\ifmmode{\mathrm{\,MeV/{\mit c}}}
          \else{MeV/{\it c}}\fi}
\def\mupp{\ifmmode{\mathrm{\,(MeV/{\mit c})^2}}
          \else{(MeV/{\it c})$^2$}\fi}

\def\gunit{\ifmmode{\,\mathrm{GeV/{\mit c}^{\,2}}}
          \else{GeV/$c^{\,2}$}\fi}
\def\pup{\ifmmode{\mathrm{\,GeV/{\mit c}}}
          \else{GeV/{\it c}}\fi}
\def\pupp{\ifmmode{\mathrm{\,(GeV/{\mit c})^2}}
          \else{(GeV/{\it c})$^2$}\fi}
\def\pum{\ifmmode{\mathrm{\,(GeV/{\mit c})^{-1}}}
          \else{(GeV/{\it c})$^{-1}$}\fi}
\def\pumm{\ifmmode{\mathrm{\,(GeV/{\mit c})^{-2}}}
          \else{(GeV/{\it c})$^{-2}$}\fi}

\def\rmunit{\ifmmode{\,\mathrm{íÜ÷/c^2}}
          \else{íÜ÷/c$^2$}\fi}
\def\rmup{\ifmmode{\mathrm{\,íÜ÷/c}}
          \else{íÜ÷/c}\fi}
\def\rmupp{\ifmmode{\mathrm{\,(íÜ÷/c)^2}}
          \else{(íÜ÷/c)$^2$}\fi}

\def\rgunit{\ifmmode{\,\mathrm{çÜ÷/c^2}}
          \else{çÜ÷/c$^2$}\fi}
\def\rpup{\ifmmode{\mathrm{\,çÜ÷/c}}
          \else{çÜ÷/c}\fi}
\def\rpupp{\ifmmode{\mathrm{\,(çÜ÷/c)^2}}
          \else{(çÜ÷/c)$^2$}\fi}
\def\rpum{\ifmmode{\mathrm{\,(çÜ÷/c)^{-1}}}
          \else{(çÜ÷/c)$^{-1}$}\fi}
\def\rpumm{\ifmmode{\mathrm{\,(çÜ÷/c)^{-2}}}
          \else{(çÜ÷/c)$^{-2}$}\fi}

\def\to {$\rightarrow$}

\def\cpc#1#2#3  {{\rm Computer Phys. Comm.}  {\bf#1}, (#3) #2}
\def\npb#1#2#3  {{\rm Nucl. Phys. B}         {\bf#1}, (#3) #2}
\def\plb#1#2#3  {{\rm Phys. Lett. B}         {\bf#1}, (#3) #2}
\def\prd#1#2#3  {{\rm Phys. Rev. D}          {\bf#1}, (#3) #2}
\def\prl#1#2#3  {{\rm Phys. Rev. Lett.}      {\bf#1}, (#3) #2}
\def\sjnp#1#2#3 {{\rm Sov. J. Nucl. Phys.}   {\bf#1}, (#3) #2}
\def\spjl#1#2#3 {{\rm Sov. JETP Lett.}       {\bf#1}, #2 (#3)}
\def\zpc#1#2#3  {{\rm Z. Phys. C}            {\bf#1}, (#3) #2}

\begin{document}
\pagestyle{empty}

\begin{flushright}
hep-ex/02 \\*[8mm]
\today
\end{flushright}
\vspace{1.8cm}

\centerline{
\parbox{12cm}{\bf
\begin{center}
\Large{The Invariant Mass Spectra Profile \\ Close to Pentaquark
         Mass Region \\ at 1.54 \gunit.
      }
\end{center}
}}
\vspace*{1.7cm}

\begin{center}
\large{
M.Zavertyaev \\
Max-Planck Institut f\"ur Kernphysik, D-69117 Heidelberg, Germany \\
P.N.Lebedev Physical Institute,117924 Moscow B-333, Russia}
\end{center}
\vspace*{1.7cm}
\hrule
\vspace*{0.5cm}

\noindent{\bf Abstract} \\
Two possible reasons for the narrow peak observation in the
invariant mass spectrum of the $p$\kzeros\ close to 1.54 \gunit\
are discussed.
\vspace*{0.5cm}
\hrule

\vfill

\clearpage

\section{Introduction}
\label{intro}

Five different experiments have reported the observation of a
narrow peak in the invariant mass spectra close to 1.54\gunit\ with a 
width below 25\munit . These peaks are considered as a signature of a
five quark resonance, the so-called  $\theta^+$.

Three photo-production experiments \cite{leps,class,saphir}
have observed the $\theta^+$ decay to $n$\kplus\ in three- or four- body 
final state. The neutron was reconstructed from the missing mass and energy.

The $\theta^+$ decay to $p$\kzeros\ was observed in 
$\nu_{\mu^-}(\overline{\nu}_{\mu^-})$ collisions with nuclei
\cite{beps}. This analysis has combined data samples from different bubble 
chamber experiments with a large range of neutrino momenta.

Using a \kplus\ beam, the same $p$\kzeros\ channel of the $\theta^+$ decay
was observed by the DIANA collaboration \cite{diana}. The Monte Carlo 
simulation
for this experiment is rather simple for the following reason: first -
the beam momentum spectrum is explicitly shown in the publication,
second - all particles in the final state ($p$ and \kzeros \to \piplus \pimin)
were reconstructed and identified.

The result of the simulation for the DIANA experiment are presented in the
section below. One section further the result of the experiment ``independent''
simulation shows the possible source of the narrow peak in the invariant
mass spectrum.

\section {Monte Carlo simulation.}
\label{mc}

\begin{figure*} 
\addtolength{\abovecaptionskip}{10pt}
\centering
\includegraphics[width=15cm]{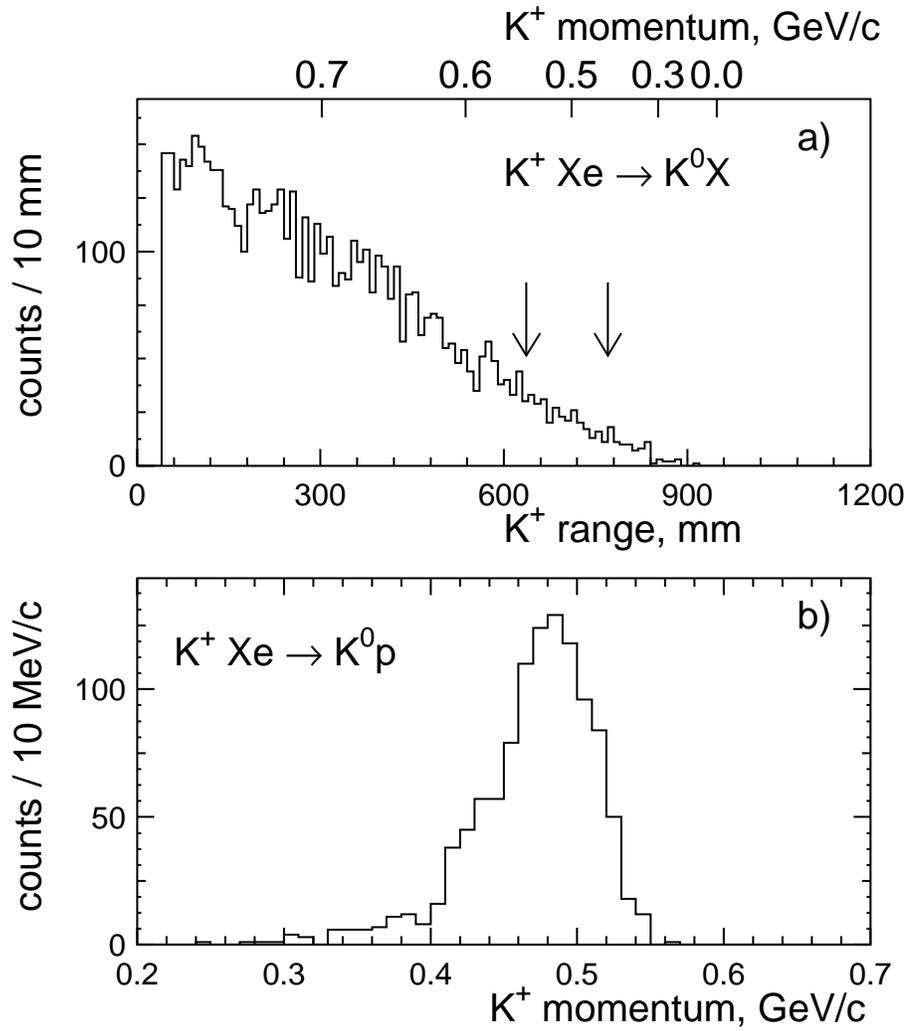}
\caption{ a) The range and equivalent momentum (upper scale) of the  
         \kplus\ beam for the events with \kzeros\ identification only.
	 The arrows indicate the momentum region of the  $p$ \kzeros\
	 selection. 
	 b) The momentum of the \kplus\ beam for the events with $p$ \kzeros\ 
	 identification. 
	 The momentum distributions were taken from  \protect\cite{diana}
	 }
\label{fig:k0xmom}
\end{figure*}

The experimental momentum spectrum of the incident beam particle \kplus\ 
(Fig.\,\ref{fig:k0xmom}) is shown for two cases: 
the momentum spectrum for all events with the identified \kzeros\ (top) ,
and for events where both the \kzeros\ and the proton were identified 
\cite{diana} (bottom). The momentum spectra look very 
different, both in momentum range and in shape. For the 
events with $p$\kzeros\ in the final state the momentum profile has an
asymmetric ``bell''-like shape with the average momentum close to 470\mup .

The \kplus momentum spectrum was simulated according to the experimental 
distribution shown in Fig.\,\ref{fig:k0xmom}b.
The momenta of the outgoing particles are generated uniformly in
the available Lorentz-invariant phase space.
Two processes are simulated :
\begin{itemize}
\item three body final state - a nucleus play a role of
      the recoil particle: \\
      \kplus Xe \to $p$\kzeros Xe'
\item two body final state - a charge-exchange reaction on
      the neutron of the Xe nucleus: \\
      \kplus $n$ \to $p$\kzeros \\
      The momentum is shared by two particles only.
\end{itemize}

\begin{figure*} 
\addtolength{\abovecaptionskip}{10pt}
\centering
\includegraphics[width=15cm]{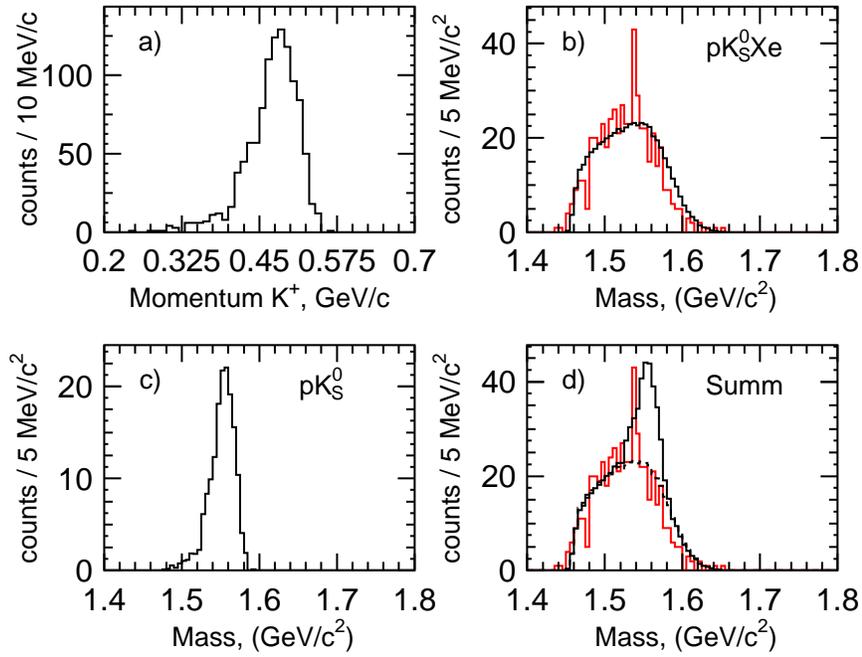}
\caption{The experimental beam momentum \protect\cite{diana}
         and MC mass spectra distribution corresponding to:
	 b) reaction \kplus Xe \to \kzeros $p Xe'$;
         c) reaction \kplus $n$ \to \kzeros $p$; 
         d) the summ of both b) and c);
	 The histogram in red corresponds to the
	 experimental mass distribution from \protect\cite{diana}.
	 }
\label{fig:massexp}
\end{figure*}

The $p$\kzeros\ invariant mass spectrum for three body final state simulation 
is shown in Fig.\,\ref{fig:massexp}b .
The comparison with the experimental distribution (``red'' histogram) shows a 
reasonable agreement in positions, widths and shapes of both spectra.

The invariant mass spectrum for the two body final state $p$\kzeros\ is shown 
in Fig.\,\ref{fig:massexp}c. In contrast to the three body final state 
$p$\kzeros Xe' the mass distribution resembles narrow peak with a sharp 
maximum at 1.55\gunit .

\begin{figure*} 
\addtolength{\abovecaptionskip}{10pt}
\centering
\includegraphics[width=15cm]{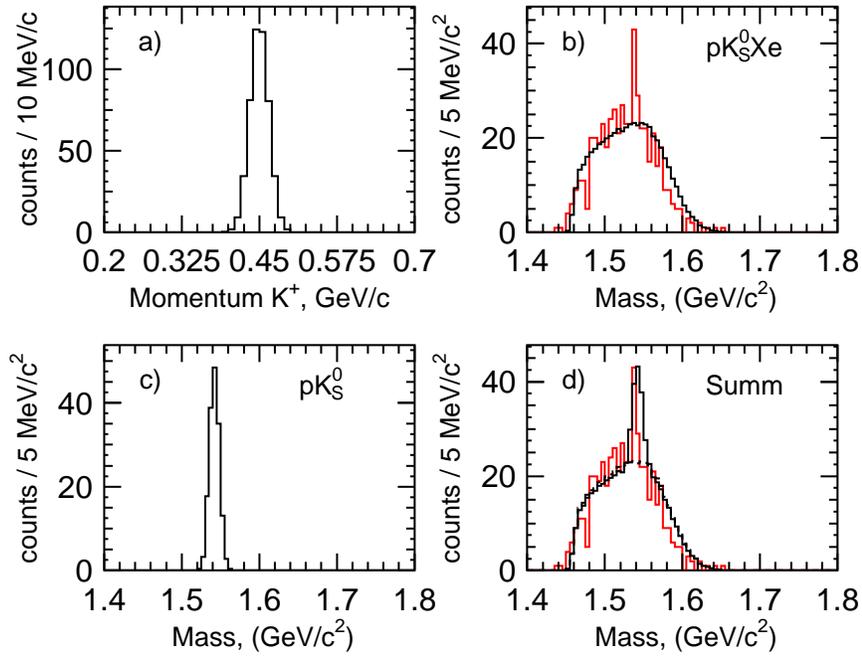}
\caption{The beam momentum simulated according to the gaussian shape
         and MC mass spectra distribution corresponding to:
	 b) reaction \kplus Xe \to \kzeros $p Xe'$;
         c) reaction \kplus $n$ \to \kzeros $p$; 
         d) the summ of both b) and c);
	 The histogram in red corresponds to the
	 experimental mass distribution from \protect\cite{diana}.
	 }
\label{fig:massmc}
\end{figure*}

The simple MC test with the \kplus\ momentum spectrum of the gaussian
shape (the mean at 450\mup\ and the width 15\mup)
shows that the narrow momentum range of the incident \kplus\ beam is the 
reason why the narrow peak is seen in the mass spectrum in case of two body
final state (Fig.\,\ref{fig:massmc}c). The shift of the \kplus\ momentum 
distribution by 20\mup\ gives the proportional shift by 10\munit\ in the mass
peak position (Fig.\,\ref{fig:massmc}d) and puts the maximum at 1.54\gunit .

\begin{figure*} 
\addtolength{\abovecaptionskip}{10pt}
\centering
\includegraphics[width=11cm]{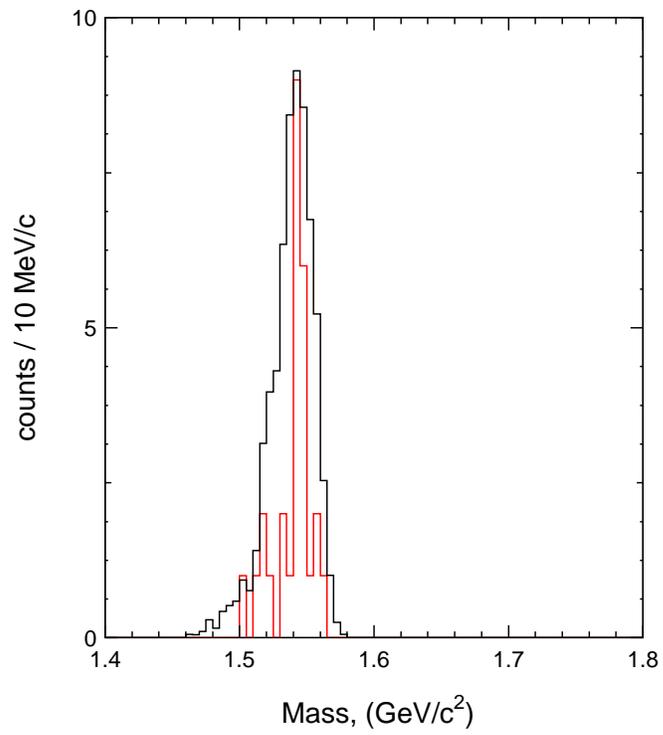}
\caption{The invariant mass distribution for p\kzeros\ combination in case,
         when at least 50\% of entries were observed in two bins (red) and
	 the ``true'' distribution shape (black) in case of big number of
	 entries. 
	 }
\label{fig:massw}
\end{figure*}

The simulation of the $p$\kzeros\ kinematic disagree with the observed
invariant mass spectrum only in one point - peaks are shifted by 
10-15\mup . This shift may come from the fact that in the MC the binding
energy of the nucleus was not included in the simulation. There is an explicit
indication in reference \cite{diana} that the \kplus\ beam momentum
was shifted by $\pm 15$\mup\ in different data sample, but there is no
indication about the corresponding statistics, accumulated with a particular
beam momenta. As result, the MC events were not weighted and that may also
cause some kind of shift in the peak position. The difference in the
width is not too important -the  MC shows that with statistic of about 30 
event, the narrow peak may be observed even if the original distribution is
wide (Fig.\,\ref{fig:massw}).  The narrow two-bin peak distribution, when
at least 50\% of entries belongs to two bins only, was
observed in 2\% cases - a non-negligible number.

\section {The kinematic ``reflection'' at 1.54\gunit .}
\label{kine}

Since the mass region around 1.54\gunit has so much importance in the narrow 
peak search, another source of trouble must be pointed out.

In a data sample of $V^0$ candidates ($V^0$ is used as a generic name
for \kzeros , \lambdazero\ or \antilambda) the narrow peak in the 
mass spectrum may be created in the following way.

 A $V^0$ decays in two particles of opposite charge. Since the type
of the particle is not known in the experiment the different mass
hypothesis have to be assigned to each track to explore all combinations.
With pion and proton/anti-proton masses assigned to corresponding tracks,
each $V^0$ may represent \kzeros , \lambdazero\ or \antilambda.

\begin{figure*} 
\addtolength{\abovecaptionskip}{10pt}
\centering
\includegraphics[width=15cm]{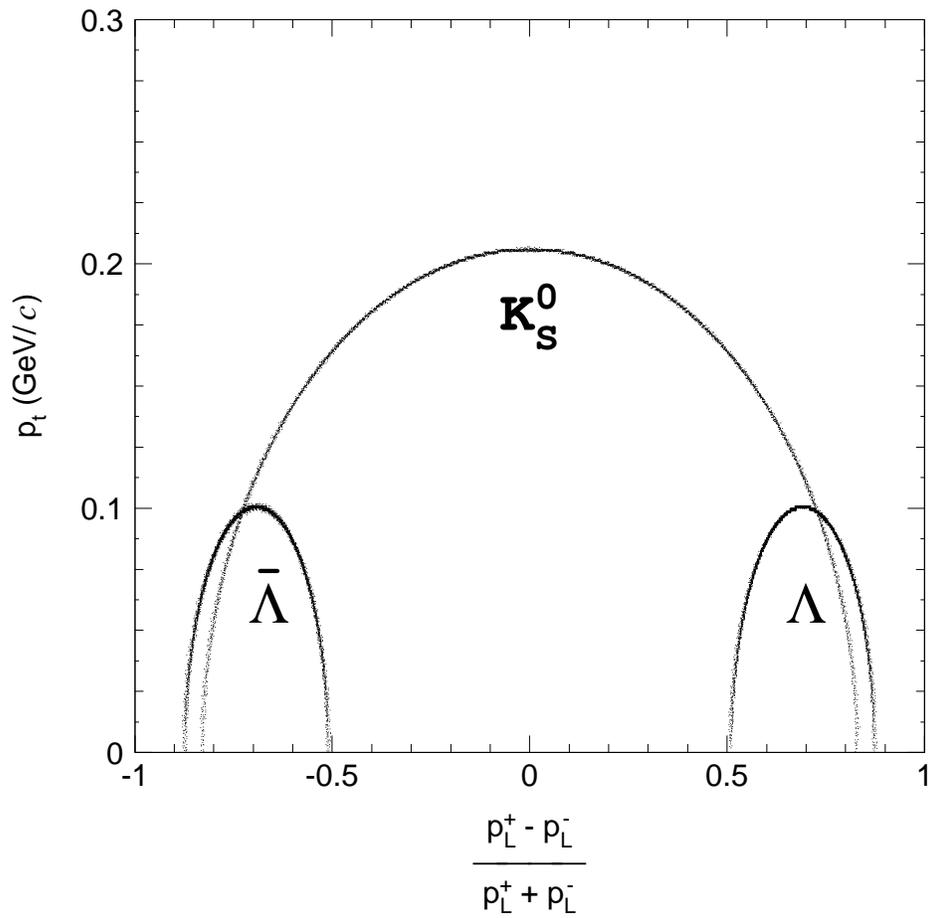}
\caption{The Armenteros-Podolansky plot for the $V^0$ candidates.
         $p^{\pm}_L$ and $p_t$ are the laboratory longitudinal and
         transverse momenta, respectively, of the decay tracks with
         respect to the $V^0$ direction. At the crossing point of the two
         ellipses the \kzeros\ and \lambdazero\ kinematically looks the same.
        }
\label{fig:armentero}
\end{figure*}

The most convenient way to present the kinematic of the $V^0$ decays 
is an  Armentero-Podolansky plot \cite{armentero}, shown in 
Fig.\,\ref{fig:armentero} , 
where the three ellipses correspond to \kzeros \to \piplus \pimin ,
\lambdazero \to $p$ \pimin\ and \antilambda \to \ap \piplus decays.
The ellipses have two overlap regions where this $V^0$ species are
kinematically indistinguishable. Therefore, without an explicit cut
on the corresponding mass hypothesis, there is always a cross
contamination. For example \kzeros\ may be not a ``true'' \kzeros\
but a \lambdazero\ or \antilambda\ in these narrow kinematic regions.

In second step the $V^0$ is then combined with
the same physical track which was already used to for the $V^0$ itself.
Two class of combination may be defined:
\begin{itemize}
\item True \kzeros\ and the \piplus\ of the \kzeros\ but considered as a 
      proton (class~1).
\item True \lambdazero\ but considered as a \kzeros\ and a positive track of
      the \lambdazero\ which is a true proton (class~2).
\end{itemize}

\begin{figure*} 
\addtolength{\abovecaptionskip}{10pt}
\centering
\includegraphics[width=11cm]{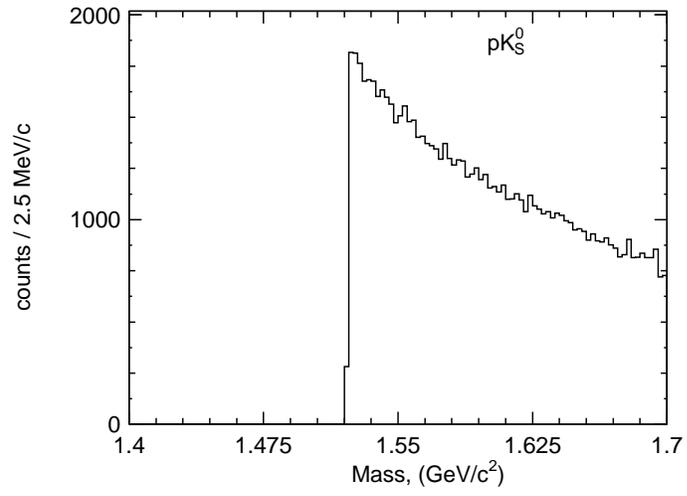}
\caption{The invariant mass distribution for p\kzeros\ combination of class~1. 
	 }
\label{fig:comb1}
\end{figure*}

\begin{figure*} 
\addtolength{\abovecaptionskip}{10pt}
\centering
\includegraphics[width=11cm]{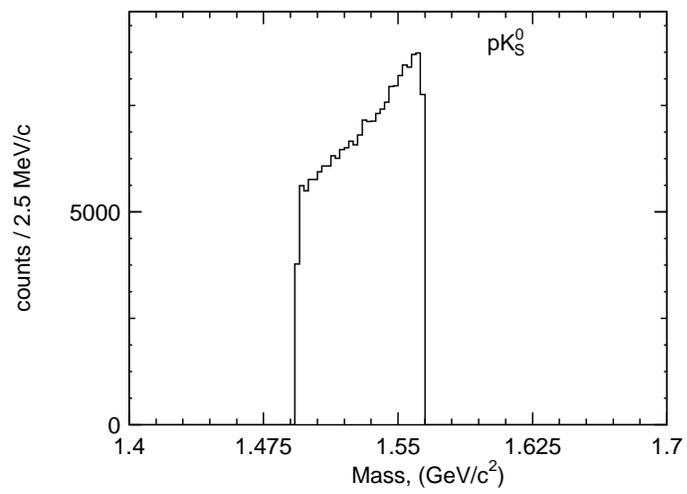}
\caption{ The invariant mass distribution for p\kzeros\ combination of class~2.
	 }
\label{fig:comb2}
\end{figure*}

Figures \ref{fig:comb1} and \ref{fig:comb2} show 
the invariant mass distributions for the two classes of combinations defined 
above.

\begin{figure*} 
\addtolength{\abovecaptionskip}{10pt}
\centering
\includegraphics[width=11cm]{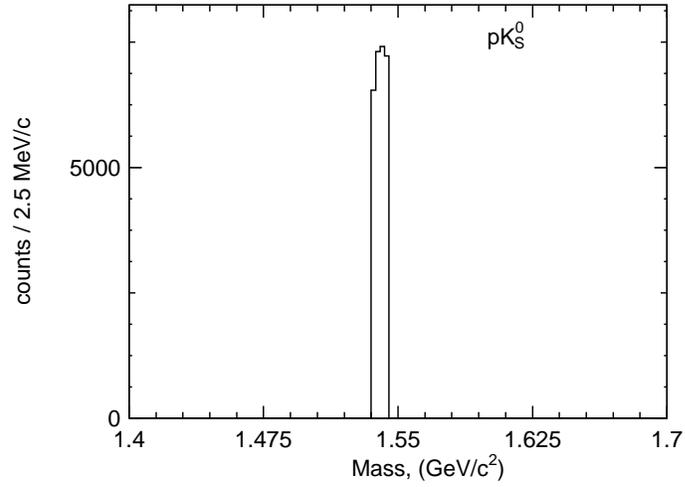}
\caption{ The invariant mass distribution for p\kzeros\ combination of class~2
          when the \kzeros\ mass is limited within the \kzeros\ mass 
          region.
	}
\label{fig:comb3}
\end{figure*}

\begin{figure*} 
\addtolength{\abovecaptionskip}{10pt}
\centering
\includegraphics[width=11cm]{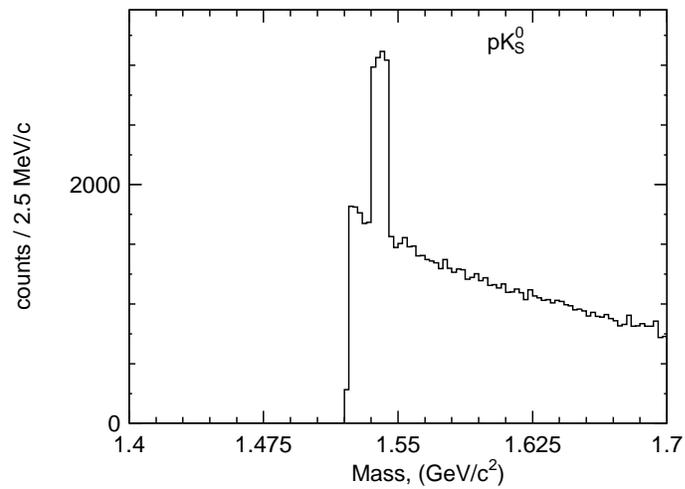}
\caption{ The combined invariant mass distribution for p\kzeros\ combination 
          of class~1 and class~2, when the \kzeros\ mass is limited within 
	  the \kzeros\ mass peak region.
	 }
\label{fig:comb4}
\end{figure*}

The standard cut used in all experimental analysis is a cut on the
invariant mass of the \piplus \pimin\ combination around \kzeros\ mass
($|m_{K^0_S} \pm m_{\pi^+\pi^-}| < 3\cdot \sigma $). The usual mass resolution
for two track combinations is limited to $\sigma $=2-4\mup .
For class~1 this cut does not change 
the invariant mass distribution (true \kzeros ), 
but for class~2 this cut removes combinations with small masses and 
retains combinations around 1.54\gunit (Fig.\,\ref{fig:comb3}). 
Fig.\,\ref{fig:comb4} shows the combined (class~1 and class~2) mass 
spectrum.

It is hardly believable that a straight forward double track counting may 
happen in a real analysis. At least for DIANA this possibility is excluded
by the nature of the experiment - in the bubble chamber all tracks are visible.
The problem might be important for the electronic experiment if the cut on
$V^0$ masses is not tight enough, thus the \kzeros\ - \lambdazero\ cross 
contamination is not removed completely, and the pool of reconstructed 
tracks is not free from ``clones'' or ``ghosts'' - de-facto the same
track reconstructed twice due to , for example, some noise in the tracking 
detectors.
The kinematic of such ``clones'' is very close to each other thus simulating
the situation discussed above, naturally with a slight variation in the mass
position and the width of the peak.

\section{Conclusion.}
\label{concl}

The result of the simple MC shows that the invariant mass spectrum of the 
$p$\kzeros combinations has a narrow structure close to 1.54\gunit\
due to the narrow momentum range of the beam. If this is true, then further
data analysis using the full moment range of the beam 
(see Fig.\,\ref{fig:k0xmom}a)
will cause the broadening of the observed peak and yield the answer about
its nature. 

The artificial peaks in the invariant mass spectra due to kinematic
reasons are well known. The only surprise is that such an artifact
precisely coincides with the value of 1.54\gunit .

The recent publication \cite{kinrefl} shows that the $nK^+$ channel is 
not free from the kinematic reflections. 

Finally the question about the pentaquark signature will be found in
a very classical way - increasing the statistic involved in the analysis
and hopefully in a different experimental conditions.

\section*{Acknowledgments}
\label{ackn}

The author acknowledge very useful discussions with J.Blouw, K.T.Kn\"opfle,
M.Schmelling and B.Schwingenheuer.

\clearpage

\end{document}